# HARNESSING SIMULTANEOUSLY THE BENEFITS OF UWB AND MBWA: A PRACTICAL SCENARIO

*Mouhamed Abdulla and Yousef R. Shayan*

Concordia University
Faculty of Engineering and Computer Science
Department of Electrical and Computer Engineering
Montréal, Québec, Canada
Email: {m_abdull, yshayan}@ece.concordia.ca

## ABSTRACT

UWB has a very large bandwidth in a WPAN network, which is best used for HD-video applications. Meanwhile, MBWA is a WMAN option optimized for wireless-IP in a fast moving vehicle. In this paper, we propose a practical engineering scenario that harnesses simultaneously the distinctive feature of both UWB and MBWA. However, this in-proximity operation of the technologies will inevitably cause mutual interference to both systems. In light of this, as a preliminary phase to coexistence, we have derived, under various circumstances, the maximum interference power limit that needs to be respected in order to ensure an acceptable system performance as requested by the new IEEE 802.20 standard.

## 1. INTRODUCTION

Without taking into account longer inter-city transportation, studies by the US Census Bureau estimate that over 100 hours a year are spent commuting to work [1]. Obviously, this indicates that a reasonable portion of a typical day is consumed while going from one place to another. And since we are in the information age, being always connected, no matter the location, is very important to users.

Because of this reason, wireless communication has steadily evolved through various so-called "revolutions" [2]. The $1^{st}$ notable revolution was in the 1990s brought by Wireless Wide Area Network (WWAN) cellular systems used for "audio" exchange. The $2^{nd}$ revolution took place around the start of the new millennium with the popularization of Wireless Local Area Network (WLAN) through the use of IEEE 802.11 for "data" transmission. Today, we are moving toward a $3^{rd}$ progression that focuses on both: wireless High Definition (HD) "video" and mobile Internet Protocol (IP) for data and Voice over IP (VoIP) applications. In particular, HD-video would be optimized for a short range Wireless Personal Area Network (WPAN); while mobile-IP would be transmitted at a substantially longer range through Wireless Metropolitan Area Network (WMAN).

In this paper, we will first discuss the possible use of HD-video through Ultra Wideband (UWB) technology. Next, we will present the purpose of the new IEEE 802.20 Mobile Broadband Wireless Access (MBWA) standard. Then, we will demonstrate a practical engineering scenario that would fully harness the principle motive of both UWB and MBWA protocols. After, we will derive a generalized interference model for the MBWA system, which under specific conditions will yield us the maximum tolerated aggregate interference power. The results acquired will eventually be used as a benchmark for future study on the coexistence of various UWB offenders toward an MBWA system.

## 2. HIGH DEFINITION VIDEO

With the recent interest in HD-video, Consumer Electronic giants such as, Hitachi, Intel, LG Electronics, Motorola, Panasonic, Samsung, Sharp, Sony, Toshiba, among others, have created interest groups to further push this initiative and to ensure interoperability. The two notable alliances are Wireless Home Digital Interface (WHDI$^{TM}$) [3] and WirelessHD$^{TM}$ [4].

In fact, WHDI operates in the 5 GHz unlicensed Industrial, Scientific and Medical (ISM) band, it supports Point-to-Multipoint (P2MP) connection, it requires No Line of Sight (NLOS), and extends to a 30 m range. On the other hand, WirelessHD operates in the 60 GHz unlicensed ISM band, with a Point-to-Point (P2P) NLOS link within a 10 m range. Both specifications are expected to be in use sometime in 2009.

Although WHDI and WirelessHD support uncompressed HD-video streaming, a third low-power alternative using lossless and low latency video compression known as wireless HD Multimedia Interface (HDMI) exists. As a matter of fact, it is currently manufactured by TZero [5] using UWB technology through the ECMA-368 standard [6]. A summary of the UWB specifications is shown in Table 1.

| | Offender | Victim |
|---|---|---|
| Technology Name | UWB | MBWA |
| Date of Approval | FCC Approval – Feb. 14, 2002 | IEEE Approval – Jun. 12, 2008 |
| Standard | ECMA 368 [for MB-OFDM] [DS-UWB-IR, TH-UWB-IR have no standards yet] | IEEE 802.20 |
| Standard Published Date | 1st Ed. – Dec. 2005 2nd Ed. – Dec. 2007 | Aug. 29, 2008 |
| Industry Consortium | WiMedia Alliance [for MB-OFDM] UWB Forum [for DS-UWB-IR] [TH-UWB-IR has no consortium] | None as of yet! [possible name: "Mobile-Fi Alliance"] |
| Network Type | WPAN | WMAN |
| Coverage Range | ≈ 10+ m [Short Range] | ≈ 15 km [Long Range] |
| MS Mobility | ≈ 3 km/hr [Pedestrian] | up to 250 km/hr [Fast Vehicular] |
| Network Topology | P2MP [up to 127 MB-OFDM devices] | P2MP |
| Base Station Needed? | No | Yes [cellular based] |
| Spectrum Licensing | Licensed Free Band | Licensed Band |
| RF – ITU Band | SHF | UHF / SHF |
| Frequency Range | 3.1 to 10.6 GHz [as per FCC Part 15] | 0.5 to 3.5 GHz |
| Channel BW | 528 MHz [for MB-OFDM – 14 channels] | 2.5 to 20 MHz [FDD-OFDMA] 5 to 40 MHz [TDD-OFDMA] 625 kHz/carrier [TDD-625k MC] |
| Peak Data Rate per User [or per carrier for 625k MC] | 110 to 480 Mbps [from 10 to 3 m range under LOS] | 1 to 4.5 Mbps [DL-OFDMA] 0.3 to 2.25 Mbps [UL-OFDMA] [values are only for the following BW: $BW_{FDD}$ = 1.25 MHz | $BW_{TDD}$ = 2.5 MHz for other BW use ratios of the values shown here] 1.493 Mbps [DL-625k MC] 0.5712 Mbps [UL-625k MC] |
| Maximum EIRP | $-41.3$ dBmW = $7.413 \cdot 10^{-8}$ W [as per FCC Part 15 with 1 MHz resolution] | 57 dBmW = 501.2 W [DL] 27 dBmW = 0.5 W [UL] |
| Dominant Fading PDF | Nakagami Fading [WLOS] | Rayleigh Fading [NLOS] |
| Duplexing Technique | TDD | FDD or TDD [OFDMA] TDD [625k MC] |
| Multiple Access and Modulation | MB-OFDM / QPSK [ECMA-368, WiMedia] DS-UWB-IR / OOK, BPSK [UWB Forum] TH-UWB-IR / PPM | OFDMA / BPSK, QPSK, QAM HC-SDMA / 625k MC [QAM] |
| Advantages | High-speed, Low Power | Mobility, Spectral Efficiency, Low Latency, Long Range |
| Disadvantages | Accurate Synchronization, Narrowband Interference, Short Range | Cost of Infrastructure |
| Potential Applications | HD-Video, Virtual Reality Gaming | Mobile-IP, VoIP |

Table 1. Specifications for UWB and MBWA standards

## 3. A NEW STANDARD: MOBILE BROADBAND WIRELESS ACCESS (IEEE 802.20)

Gradually, 4th Generation (4G) systems are in the making and are expected to dominate the marketplace in the years to come. Specifically, there seems to be two general options available as we move toward 4G: either upgrade available 3 and 3.5G standards or simply use a new technology [7]. On the surface, some may consider an upgrade as a cost-effective option because it requires no or minor infrastructure modification. However, an upgrade would always be constraint to backward compatibility issues, and this would result in a suboptimal system which often defeats the purpose.

Due to this, in June 2008, a new 4G cellular standard called IEEE 802.20 or MBWA was approved by the IEEE Standard Association Board [8]. Soon after, in August 2008, the first active MBWA standard was published and released to the public [9]. In essence, the aim of this protocol is to fill the current demand gap of increased mobility of up to 250 km/hr and a spectral efficiency of at least 1 b/s/Hz/cell [10].

Figure 1, shows how this protocol connects and compares to other channels and systems [11]. As it can be seen, MBWA is basically the "missing link" between available WMAN and WWAN standards. Notably, on one hand, WMAN's WiMAX has a high Bandwidth (BW) that could theoretically reach 70 Mbps. But perhaps a more practical data rate for the Mobile WiMAX is 10 Mbps over 2 km under NLOS. While high throughput is true, IEEE 802.16e-2005 can only support a Mobile Station (MS) with vehicular speed of 60+ km/hr [12]. On the other hand, the mobility provided by a cellular system is quite substantial. Therefore, it was natural to combine these advantages to form what is now known as IEEE 802.20 technology.

Pursuing this further, when compared to other mobile system such as: EDGE, UMTS, CDMA2000 1xRTT and 1xEV, MBWA has the highest spectral efficiency. And this is always needed because it will use the licensed channel bandwidth more adequately, and hence result in a cost effective approach to providers and consumers alike.

In addition to exceptional mobility and spectral efficiency, MBWA is specifically optimized for mobile-IP connectivity of data. Furthermore, because the system has low latency with a frame Round Trip Time (RTT) of at most 10 ms [10], MBWA may also be used for telephony applications such as VoIP. In fact, there is a direct relation between latency and performance which may be traded among each other [13] for more enhanced real-time applications and to satisfy Quality of Service (QoS).

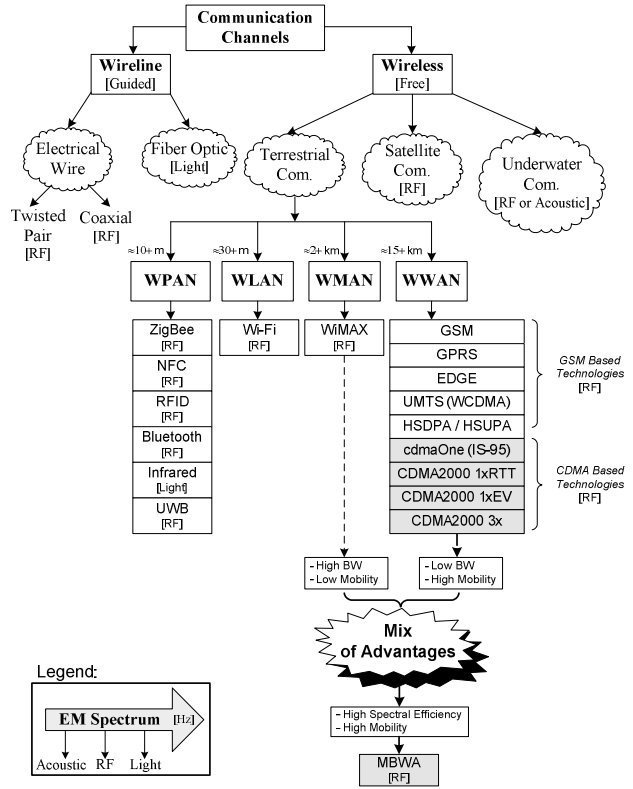

Figure 1. Comparing MBWA to other systems

As for the specification, this standard defines only the lower Physical (PHY) and Media Access Control (MAC) layers of the Open Systems Interconnection (OSI) model. And often, these two partitions are jointly referred to as the "Air Interface Layer". Also, in order to enable vast compatibility with diverse systems, the upper network layers are not specified by the 802.20 protocol.

Furthermore, the MBWA air interface layer allows two possible modes of operations: "wideband OFDMA" mode and "625 kHz multicarrier" (625k-MC) mode. The wideband mode enables Time Division Duplex (TDD) or Frequency Division Duplex (FDD) for Uplink (UL) and Downlink (DL); whereas the 625k-MC mode only supports TDD duplexing. It is worth mentioning that according to the standard a Base Station (BS) and a corresponding MS may opt to carry only one of the modes, and need not to have both capabilities simultaneously active.

Undoubtedly, this new standard has many benefits. Nonetheless, because it is a novel technology, infrastructure cost may be required; even though it is expected to be able to reuse existing BSs [10]. A summary of the MBWA specifications is shown in Table 1.

## 4. POTENTIAL APPLICATION FOR UWB & MBWA

To recap, in the previous sections, we talked about the current application interest by industry researchers in wireless-HD and mobile-IP through WPAN/UWB and WMAN/MBWA technologies respectively. Here, we will give a practical example that would exploit the purpose of each of the standards mentioned while operating at the same time.

In this monograph, we selected MBWA as the leading 4G technology of the future. And as mentioned above some of the key features of this protocol are: mobility, spectral efficiency, low latency optimized for IP services. It seems that among the qualities of the standard listed, mobility makes this system unique. In general, mobility is categorized into 6 classes [14], as explained with examples in Table 2.

| Mobility Type | Velocity | System Example |
|---|---|---|
| Stationary | 0 km/hr | Fixed WiMAX |
| Pedestrian | up to 10 km/hr | UWB |
| Vehicular | up to 100 km/hr | Mobile WiMAX |
| High-speed Vehicular | up to 500 km/hr | MBWA |
| Aeronautical | up to 1,500 km/hr | NAVAIDS |
| Satellite | up to 27,000 km/hr | GPS |

Table 2. Mobility classification for telecommunications

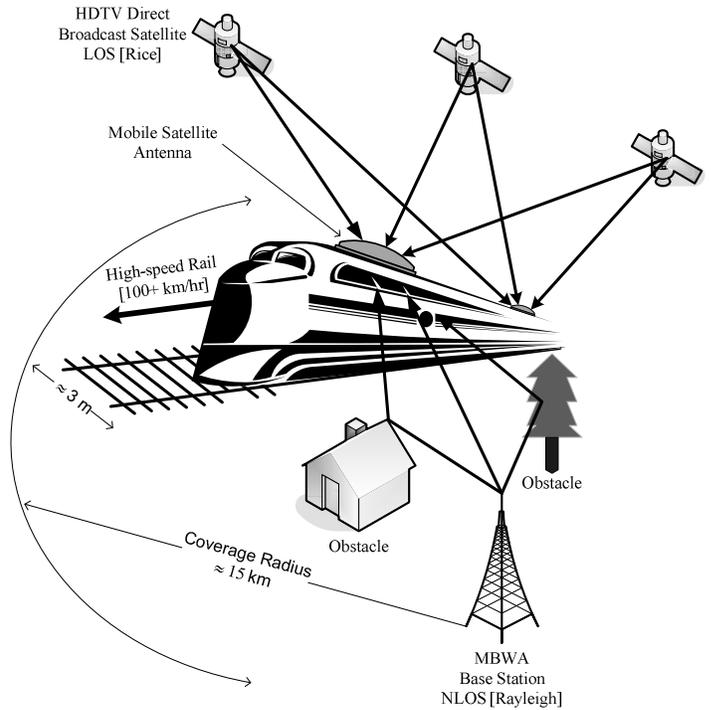

Figure 2. A practical scenario harnessing simultaneously UWB and MBWA wireless technologies

And to emphasize, IEEE 802.20, has designated that an MS could go at most 250 km/hr. In this situation we may consider a car, a coach, a recreational vehicle or maybe a train. In fact the train might be the best harnessing option due to its supported velocity. Besides, several nations have already in operation such high-speed rail.

On a parallel note, while considering WPAN, we noted that among available standards, UWB has distinctive qualities such as: low power requirements, high throughput of at most 480 Mbps, and requires no license for operation [11]. It turns out, that one of the best applications that would take advantage of its broad BW would be to transmit wirelessly HD-video.

Now that we have discretely selected a real-life practical application that would harness fully each of the standards, how could we merge these two ideas together? Figure 2, describes visually a practical scenario that would use both technologies.

Notice that several HDTV direct broadcast satellites transmit live DVD-quality video, which are then received by one or several mobile satellite antennas located on the roof of the train [15]. In fact the idea of receiving live TV on a fast moving vehicle is being slowly implemented by several rail organizations. Also, private companies such as TracVision [16] are selling the mobile antenna and the service to consumers to be placed on cars, coaches, and even boats.

It should be noted that the setup of Figure 2, does carry challenges that require further research. For instance, how well, will a Receiver (Rx) detect a HD signal while moving at a very high-speed? Also, in the case where the train would pass through a tunnel and no direct LOS to the satellite exists, what should be done to continue the live streaming of the HD-video? Should a local BS that requires NLOS be used as a relay until direct link to satellite is once again established? And if so, how would this affect latency, handoff, etc? Nevertheless, for the moment we could ignore these hurdles to further proceed in the analysis.

Going back to the scenario, once the HD-video signal is received by the mobile antenna, it then goes through a wire to a satellite-box within a railway car as shown in Figure 3. Next, the box transmits wirelessly the video signal to in-car UWB users through a master UWB system. Notice, that we strategically placed the master system in the middle of the roof to have the best coverage. And clearly, without going into details, the master device should use a non-isotropic antenna maximizing the car space coverage and minimizing useless back-lobe radiation.

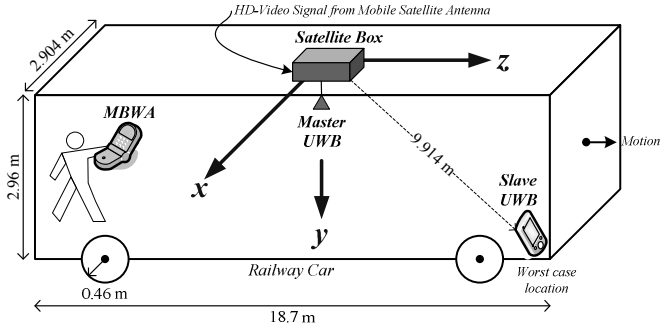

Figure 3. Railway car using TGV dimensions

Admittedly, because of the high popularity of the French high-speed train, we decided to use the dimensions of the TGV [17]. In such a configuration, using vector geometry, with the help of the defined coordinate system, we found that the farthest worst case distance for a UWB slave is about 10 m away from the master. And, as we known from Table 1, UWB with 10 m separation delivers a throughput of about 110 Mbps; which is still tolerable for video applications. Essentially, this indicates that we could consider each railway car as a WPAN space.

## 5. INTERFERENCE MODEL

Now that the stage is set for a practical application that harnesses to the extreme the benefits of UWB and MBWA, it would be critical to study the coexistence of these systems. As a preliminary phase toward coexistence, we will start by looking at the interference power seen by an MBWA victim. To proceed in the analysis, we may utilize Figure 4, which depicts an MBWA-Rx for the MS.

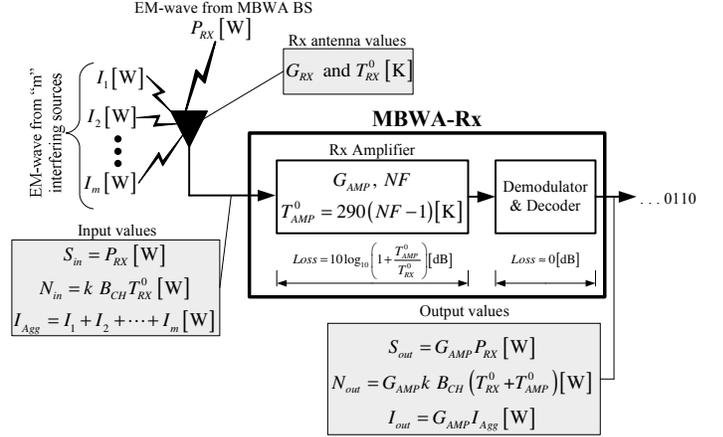

Figure 4. Model for mobile station MBWA receiver

While ignoring decibel (dB) loss from the demodulator and the decoder, we obtain the following Signal to Noise Ratio (SNR) measured at the output of the Rx:

$$SNR_{out} = \frac{S_{out}}{N_{out}} = \frac{P_{RX}}{k \, B_{CH} \left( T_{RX}^0 + T_{AMP}^0 \right)} \quad (1)$$

Proceeding in a similar fashion, we get the Signal to Interference plus Noise Ratio (SINR):

$$SINR_{out} = \frac{S_{out}}{\left( I_{out} + N_{out} \right)} = \frac{P_{RX}}{\left( I_{Agg} + k \, B_{CH} \left( T_{RX}^0 + T_{AMP}^0 \right) \right)} \quad (2)$$

In general, wireless protocols often specify the maximum allowed degradation by a system in dBs. And in literature such as [18] and [19], among others, a definition for the degradation in linear notation is given as:

$$d \triangleq \frac{SNR_{out}}{SINR_{out}} \quad (3)$$

Substituting (1) and (2) into (3), we obtain:

$$d = \frac{SNR_{out}}{SINR_{out}} = \frac{\left( I_{Agg} + k \, B_{CH} \left( T_{RX}^0 + T_{AMP}^0 \right) \right)}{k \, B_{CH} \left( T_{RX}^0 + T_{AMP}^0 \right)} = \frac{\left( I_{Agg} + N \right)}{N} \quad (4)$$

Where $I_{Agg}$ is the aggregate interference power seen by an MBWA antenna; and for the simplest case, $N$ is the thermal noise power measured at the output of the MS Rx.

At this stage, from (4), we could obtain the degradation in decibels and then isolate for the interference to find:

$$I_{Agg} = N \left( 10^{d[dB]/10} - 1 \right) [W] \quad (5)$$

More elegantly the interference power in dBmW becomes:

$$I_{Agg}[dBmW] = N[dBmW] + 10\log_{10}\left(10^{d[dB]/10} - 1\right) \quad (6)$$

Where the noise power is:

$$N = k\, B_{CH} \left\{ T_{RX}^0 + 290\left(10^{NF[dB]/10} - 1\right) \right\} \text{ [W]} \quad (7)$$

Or in dBmW it becomes:

$$N[dBmW] = -198.6 + 10\log_{10}\left(B_{CH}\left\{T_{RX}^0 + 290\left(10^{NF[dB]/10} - 1\right)\right\}\right) \quad (8)$$

So overall, the aggregate interference power seen by an MBWA Rx is a function of the system degradation, the antenna temperature, the Noise Figure (NF), and the channel BW:

$$I_{Agg} = f\left(d\,;\, T_{RX}^0\,;\, NF\,;\, B_{CH}\right) \quad (9)$$

The MBWA air interface layer standard [9], and the 802.20 permanent evaluation criteria document [20] have specified the values needed in (9) as listed in Table 3.

|  | Base Station | Mobile Station |
|---|---|---|
| $d_{max}$ | 0.5 dB | 0.5 dB |
| $T_{RX}$ | 290 K° | 290 K° |
| NF | 5 dB | 10 dB |

Table 3. Degradation, Rx temperature, and NF for MBWA

After applying the values for the MS, and adding 60 dB because $B_{CH}$ will always be in MHz, the model reduces to:

$$I_{Agg}[dBmW] = -113.112 + 10\log_{10}(B_{CH}) \quad (10)$$

At present, we should be careful in the analysis because IEEE 802.20 standard supports mobility. And, as we known, the faster the MS travels (m/s), the slower the throughput (b/s) will be. In the MBWA system requirement document [14], the minimum spectral efficiency $\eta$ is given as a function of the MS speed and the link:

$$\eta = \eta(mobility\,;\, link) \text{ [bps/Hz]} \quad (11)$$

A summary is shown in Table 4 for pedestrian and high-speed vehicular mobility.

When we first introduced MBWA, we mentioned that two modes of operations exist: wideband OFDMA and 625k-MC. The later, uses a channel BW of 625 kHz per carrier for DL and UL; where a single user may utilize multiple carriers [13]. Therefore, while assuming a constant 625 kHz BW and a varying data rate, using (10), we get the aggregate interference per carrier: -115.15 dBmW.

| MS Mobility | 3 km/hr | | 120 km/hr | |
|---|---|---|---|---|
| Link | DL | UL | DL | UL |
| Spectral Efficiency $\eta$ [bps/Hz] | 2.0 | 1.0 | 1.5 | 0.75 |

Table 4. Dependencies of the spectral efficiency

On the other hand, the OFDMA mode supports BW from 2.5 to 20 MHz for FDD and 5 to 40 MHz for TDD [9]. Using the peak throughput per user from Table 1, we may derive the data rate for various possible BW as shown in Table 5.

| | $B_{CH-FDD}$ [MHz] | 2.5 | 5 | 7.5 | 10 | 12.5 | 15 | 17.5 | 20 |
|---|---|---|---|---|---|---|---|---|---|
| | $B_{CH-TDD}$ [MHz] | 5 | 10 | 15 | 20 | 25 | 30 | 35 | 40 |
| DL | $R_{Peak-Low}$ [Mbps] | 2 | 4 | 6 | 8 | 10 | 12 | 14 | 16 |
| DL | $R_{Peak-High}$ [Mbps] | 9 | 18 | 27 | 36 | 45 | 54 | 63 | 72 |
| UL | $R_{Peak-Low}$ [Mbps] | 0.6 | 1.2 | 1.8 | 2.4 | 3 | 3.6 | 4.2 | 4.8 |
| UL | $R_{Peak-High}$ [Mbps] | 4.5 | 9 | 13.5 | 18 | 22.5 | 27 | 31.5 | 36 |

Table 5. Peak data rate per user under various conditions

As it should be obvious from the table above, the throughout is a function of the channel BW, the link, and the side (lower or higher) of the peak rate:

$$R_b = R_b\left(B_{CH}\,;\, link\,;\, side\right) \text{ [bps]} \quad (12)$$

And the spectral efficiency is defined as:

$$\eta = \frac{R_b}{B_{CH}} \text{ [bps/Hz]} \quad (13)$$

If we solve for the channel BW in (13) and then substitute (11) and (12) we get:

$$B_{CH} = \frac{R_b}{\eta} = \frac{R_b\left(B_{CH}\,;\, link\,;\, side\right)}{\eta(mobility\,;\, link)} \text{ [Hz]} \quad (14)$$

Now, after plugging (14) into (10), the interference becomes:

$$I_{Agg}[dBmW] = -113.112 + 10\log_{10}\left(\frac{R_b\left(B_{CH}\,;\, link\,;\, side\right)}{\eta(mobility\,;\, link)}\right) \quad (15)$$

Using (15) alongside Tables 4 & 5, we obtain the maximum aggregate interference power tolerated by an MBWA MS under various conditions. Figures 5 & 6 show graphs of the obtained results. Also, we plotted the interference for both extremities of the data rate so as to easily perceive the DL and UL diversity in regard to the maximum threshold. Surprisingly, while using the parameters of the standard, we can notice that high mobility allows on average 1.33 mW of extra interference as compared to a pedestrian MS.

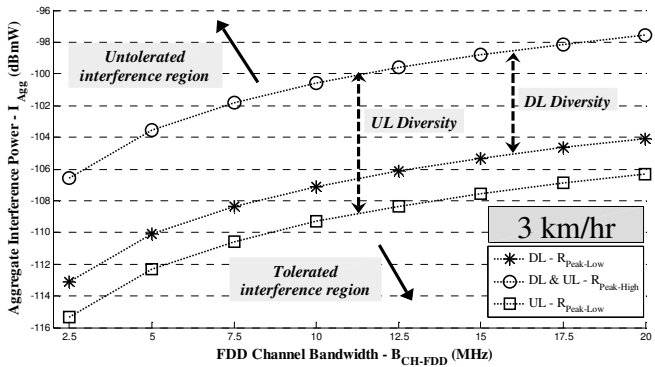

Figure 5. Max. interference power for pedestrian mobility

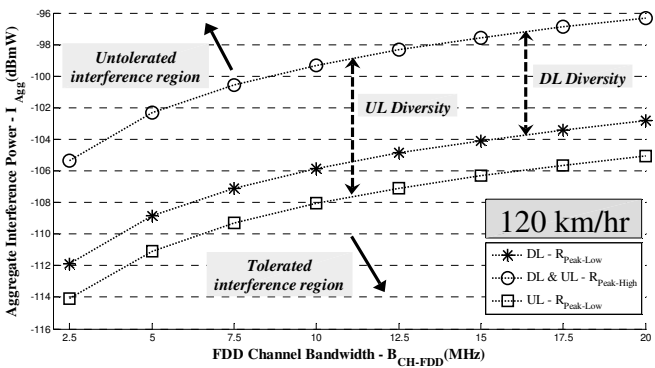

Figure 6. Max. interference power for high-speed mobility

## 6. CONCLUSION

In this paper, we discussed two practical applications: wireless-HD, and mobile-IP. We then showed a harnessing scenario that attempts to fully exploit the features of UWB and MBWA. Following this, we derived an expression for the maximum aggregate interference power allowed under diverse conditions. The result obtained will be used in future research as a benchmark to coexistence analysis.

## 7. REFERENCES

[1] Commuting to work; US Census Bureau. http://www.census.gov/Press-Release/www/releases/archives/american_community_survey_acs/001695.html

[2] B. Bing, "Broadband Wireless Access - The Next Wireless Revolution," *Proceedings of the 4th Annual Communication Networks and Services Research Conference (CNSR 2006),* pp. 14, 2006.

[3] WHDI[TM] (5 GHz). http://www.whdi.org

[4] WirelessHD[TM] (60 GHz). http://www.wirelesshd.org

[5] TZero Technologies (UWB). http://www.tzerotech.com

[6] "ECMA-368: High Rate Ultra Wideband PHY and MAC Standard, 2nd ed." Dec. 2007.

[7] G. Lawton, "What lies ahead for cellular technology?" *Computer,* vol. 38, pp. 14-17, 2005.

[8] IEEE Approves Standard for MBWA. http://standards.ieee.org/announcements/802.20approval.html

[9] "IEEE Standard for Local and metropolitan area networks Part 20: Air Interface for Mobile Broadband Wireless Access Systems Supporting Vehicular Mobility - Physical and Media Access Control Layer Specification," *IEEE Std. 802. 20-2008,* pp. 1-1039, 2008.

[10] W. Bolton, Yang Xiao and M. Guizani, "IEEE 802.20: mobile broadband wireless access," *IEEE Wireless Communications,* vol. 14, pp. 84-95, 2007.

[11] Y. R. Shayan and M. Abdulla, "Terrestrial Wireless Communications," *NATO Research and Technology Organization (RTO), Applied Vehicle Technology-Health Management for Munitions Task Group (AVT-160)*, CD-ROM, Montréal, Québec, Canada, October 14-16, 2008.

[12] WiMAX Forum, "Mobile WiMAX: A Performance and Comparative Summary".

[13] A. Greenspan, M. Klerer, J. Tomcik, R. Canchi and J. Wilson, "IEEE 802.20: Mobile Broadband Wireless Access for the Twenty-First Century," *IEEE Communications Magazine,* vol. 46, pp. 56-63, 2008.

[14] IEEE 802.20 WG, "System Requirements for IEEE 802.20 Mobile Broadband Wireless Access Systems – Version 14," IEEE 802.20 PD-06.

[15] "The Mobile Antenna," *Satellite Evolution Group,* pp. 28-32, 2007.

[16] TracVision. http://www.kvh.com/tracvision_kvh

[17] TGV dimensions of a typical trainset. http://www.railfaneurope.net/tgv/dimensions.html

[18] R. Giuliano, F. Mazzenga and F. Vatalaro, "On the interference between UMTS and UWB systems," *IEEE Conference on Ultra Wideband Systems and Technologies (UWBST 2003)*, pp. 339-343, 2003.

[19] Christian Politano et al. "Regulation and standardization" in *UWB Communication Systems : A Comprehensive Overview.* , vol. 5, New York, NY: Hindawi Publishing Corporation, 2006, pp.471-492.

[20] IEEE 802.20 WG, "802.20 Permanent Document, Evaluation Criteria – Version 1.0," IEEE 802.20 PD-09.